\documentclass[pra,aps,showpacs,twocolumn,a4paper,tightenlines,balancelastpage]{revtex4}
\usepackage{amsmath,amsfonts,amssymb,mathrsfs,bm}
\usepackage{verbatim,psfrag,times,CJK}
\usepackage{graphicx,graphics,color,epsfig}
\usepackage{flafter}

\usepackage{indentfirst}
\begin{document}

\newcommand{\ket}[1]{| #1 \rangle}
\newcommand{\bra}[1]{\langle #1 |}

\title{Deterministic creation of stationary entangled states by dissipation}

\author{A. F. Alharbi}
\author{Z. Ficek}
\email{zficek@kacst.edu.sa}
\affiliation{The National Centre for Mathematics and Physics, KACST, Riyadh 11442, P.O. Box 6086, Saudi Arabia}

\date{\today}

\begin{abstract}
We propose a practical physical system for creation of a stationary entanglement by dissipation without employing the environment engineering techniques. The system proposed is composed of two perfectly distinguishable atoms, through their significantly different transition frequencies, with only one atom addressed by an external laser field. We show that the arrangement would easily be realized in practice by trapping the atoms at the distance equal to the quarter-wavelength of a standing-wave laser field and locating one of the atoms at a node and the other at the successive antinode of the wave. The undesirable dipole-dipole interaction between the atoms, that could be large at this small distance, is adjusted to zero by a specific initial preparation of the atoms or by a specific polarization of the atomic dipole moments. Following this arrangement, we show that the dissipative relaxation can create a stationary entanglement on demand by tuning the Rabi frequency of the laser field to the difference between the atomic transition frequencies. The laser field dresses the atom and we identify that the entangled state occurs when the frequency of one of the Rabi sidebands of the driven atom tunes to frequency of the undriven atom. It is also found that this system behaves as a cascade open system where the fluorescence from the dressed atom drives the other atom with no feedback. 
\end{abstract}


\maketitle

Engineering a long-lived entanglement between atoms or ions located at small distances and interacting with a vacuum reservoir is a challenging problem in the quantum information science~\cite{lo05,mm07,kc09,hc09,hm10}.  The problem has already been the subject of several papers but attention was concentrated mainly on the effect of the direct interatomic dipole-dipole coupling and on the role of dark states for that entanglement~\cite{ph99,bb00,b02,j02,ft03,nn04,mo07,ft08,lz09,zl09,ha09,ns09,dj10,ad10,me10}. Although closely spaced atoms can occupy an area of small sizes, which has the advantage of controlling and manipulating the dissipative relaxation in the system, it is practically difficult to keep the closely spaced atoms at fixed positions. At small distances the dipole-dipole interaction is strong causing the atoms to move or rapidly oscillate around their equilibrium positions~\cite{m02}. Therefore, many schemes have been proposed for entangling atoms or atomic ensembles separated by a large distance, much larger than an optical wavelength~\cite{zg00,os01,xl05,nf07,gv10,km10,mp10}. In this case, the direct dipole-dipole interaction between the atoms can be ignored, but at the cost of entanglement stability due to difficulties in controlling the dissipative relaxation of atoms spread over large regions. The relaxation is a source of decoherence that leads to irreversible loss of information encoded in the internal states of the atoms and thus is regarded as the main obstacle in the generation of persistent entanglement. In contrast, an idea of "quantum-state engineering" by dissipation have been invented, where the dissipative relaxation can be used to drive atoms to a long-lived entangled state~\cite{dm08,kb08,bz10,dt10,vw09,bs10}. 

In the following we consider a simple physical system where a stationary entanglement can be created  by a dissipative exchange of photons between two separate systems coupled to a decohering environment. The proposed system is composed of two non-identical atoms which decay in a vacuum reservoir with no coherent couplings and no hidden initially populated entangled states present. The atoms are placed at a small but finite distance $r_{12}$ and are driven by a standing-wave laser field propagating in the direction parallel to the inter-atomic axis and are also coupled to the rest of the multi-mode radiation field which is in an ordinary vacuum state. Each atom is modeled as a two-level system (qubit) with the ground state $\ket{g_i}$ and the excited state  $|e_i\rangle$, $(i=1,2)$ separated by a transition frequency $\omega_{i}$. We assume that in the absence of the laser field the atoms are perfectly distinguishable that the frequency difference $\Delta_{0}=\omega_{1}-\omega_{2}$ is quite large to prevent photons emitted by each atom to induce changes in the other atom. In addition, we assume that the atomic dipole moments $\vec{\mu}_{i}$ are not necessary equal but are parallel to each other, and arrange the external driving of the system such that only one atom is coupled to the laser field. This coupling could be achieved in practice by placing one of the atoms at a node and the other at the successive antinode of the standing wave. With this arrangement, we shall assume that only atom~$2$ is addressed by the laser field. 

The dynamics of the system is described by the density operator $\rho$, which in the interaction picture satisfies the master equation
\begin{eqnarray}
\dot{\rho} = -\frac{i}{\hbar} \left[H_{0} + H_{d},\rho\right] +{\cal L}(\rho) ,\label{a1}
\end{eqnarray} 
where 
\begin{eqnarray}
H_{0} = \hbar (\Delta_{0}\!+\!\Delta_L)  S_1^z + \hbar \Delta_L  S_2^z 
+ \frac{1}{2} \hbar\Omega_{0}\!\left(S^+_{2}\!+\!S_{2}^{-}\right) ,\label{a2}
\end{eqnarray}
is the Hamiltonian of the atoms and the interaction of the atom~$2$ with the laser field,
\begin{eqnarray}
H_{d} = \hbar \Omega _{12}\left(S_{1}^{+}S_{2}^{-} +{\rm H.c.}\right)  \label{a3} 
\end{eqnarray}
is the coherent dipole-dipole interaction between the atoms with $\Omega_{12} = {\rm Re}U_{12}$, and the Liouvillian ${\cal L}(\rho)$ describing the dissipative relaxation of the system is of the form~\cite{leh,ft02}
\begin{eqnarray}
{\cal L}(\rho) = -\sum _{i,j=1}^{2}\gamma_{ij}\!\left(\left[S_i^+,S_{j}^-\rho\right]\!+\!\left[\rho S_{i}^+,S_{j}^{-}\right]\right)  \label{a4}
\end{eqnarray}
with $\gamma_{12} = \gamma_{21} = -{\rm Im}U_{12}$, and 
\begin{eqnarray}
U_{12} &=& -\frac{3}{2}\sqrt{\gamma_{1}\gamma_{2}}\left\{ \left(1-\cos^2 \eta\right)\frac {1}{k r_{12}}\right. \nonumber\\
&+&\left. \left(1\!-\!3\cos^{2}\!\eta\right)\!\left[\frac{i}{(kr_{12})^2}-\frac{1}{(kr_{12})^3}\right]\!\right\}{\rm e}^{ikr_{12}} . \label{a5}
\end{eqnarray}
Here, $S^{+}_{i} (S^{-}_{i})$ is the raising (lowering) operator of the $i$th atom and $S^{z}_{i}$ describes its energy, $\Omega_{0}$ is the resonant Rabi frequency of the laser field at the position of the driven atom, $\Delta_L=\omega_2-\omega_L$ is the detuning of the transition frequency of the driven atom from the laser field frequency $\omega_{L}$, and $2\gamma_{ii}\equiv 2\gamma_{i}$ is the relaxation  rate of the~$i$th atom.

The parameters $\Omega_{12}$ and $\gamma_{12}$, symmetrical with respect to the exchange of the atoms $(1\leftrightarrow 2)$ even if the atoms are not identical, describe mutual interaction of the atoms through the coupling to the common reservoir. They depend on the distance $r_{12}$ between the atoms and the orientation $\eta$ of the atomic dipole moments in respect to the interatomic axis. The parameters are of importance only in systems of atoms separated by small or intermediate distances relative to the radiation wavelength, $kr_{12}\leq 1$, and vanish at $kr_{12}\gg 1$. However, they can also vanish, but not simultaneously at some finite distances. We shall use this fact to eliminate the dipole-dipole interaction $\Omega_{12}$ leaving only the dissipative interaction $\gamma_{12}$. It is easily verified that $\Omega_{12}$ vanishes at the distance $r_{12} =\lambda/4$ and for a fixed $\eta = 0.304 \pi$.
In practical terms, this could correspond to atoms trapped at fixed positions, the Raman-Nath approximation, and the atomic dipole moments polarized at a fixed direction. 
This is compatible with many experiments on cooling of trapped atoms, where the storage time of the trapped atoms is long, so that they are essentially motionless and lie at known and controllable distances from one another~\cite{lo05,mm07,kc09,hc09,hm10,m02}.
An alternative way is to identically prepare the atoms with no information about the orientation of their dipole moments. In this case, $U_{12}$ should be averaged over all orientations of the dipole moments, so that $\cos^{2}\eta = 1/3$ and then $\Omega_{12}=0$ at $r_{12}=\lambda/4$.
The fact that $\Omega_{12}$ can be zero at $r_{12}=\lambda/4$ would allow for an experimentally difficult manipulation of atoms trapped at small distances. 

We first perform numerical analysis of entangled properties of the system by solving the master equation (\ref{a1}) for the steady-state, $\dot{\rho}=0$, using the quantum optics toolbox for matlab~\cite{smt1}. We quantify entanglement in terms of the concurrence that relates entangled properties to the coherence properties of the atoms~\cite{woo}. 

\begin{figure}[htpb!]
 \includegraphics[width=0.8\columnwidth]{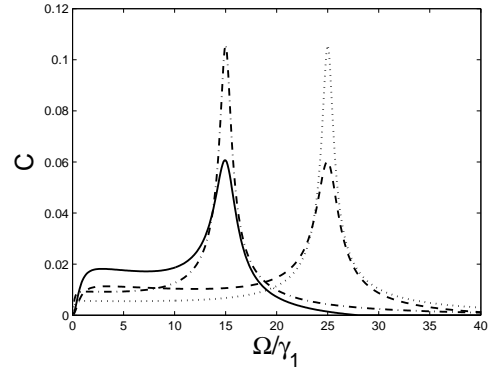}
\caption{{\footnotesize The steady-state concurrence versus the normalized Rabi frequency for $\Delta_{L}=0 \, (\Omega =\Omega_{0})$ and different $\Delta_{0}$ and $\gamma_{2}/\gamma_{1}$: $\Delta_{0}=15\gamma_{1}, \gamma_{2}/\gamma_{1} =1$ (solid line), $\Delta_{0}=15\gamma_{1}, \gamma_{2}/\gamma_{1}=5$ (dashed line), $\Delta_{0}=25\gamma_{1}, \gamma_{2}/\gamma_{1}=1$ (long dashed line) and $\Delta_{0}=25\gamma_{1}, \gamma_{2}/\gamma_{1}=5$ (dash-doted line).}}
\label{fig1}
\end{figure}

Figure~\ref{fig1} shows numerical results for the concurrence plotted as a function of the Rabi frequency~$\Omega_{0}$ for the case of resonant driving, $\Delta_{L}=0$ and different $\Delta_{0}$ and also different ratios of the atomic relaxation rates. We see that the concurrence is insensitive to $\Omega_{0}$ until $\Omega_{0} =\Delta_{0}$. At this specific value of the Rabi frequency, the concurrence exhibits a sharp peak whose amplitude is independent of $\Delta_{0}$. Consequently, the amount to which the atoms are entangled by dissipation is independent of the frequency difference $\Delta_{0}$. Another interesting, and perhaps surprising feature seen in Fig.~\ref{fig1} is the dependence of the amplitude of the concurrence on the ratio $\gamma_{2}/\gamma_{1}$, that the seemingly symmetric system in respect to the rate of exchanging an excitation, i.e. $\gamma_{12}=\gamma_{21}$, exhibits an asymmetry in the dependence of the concurrence on $\gamma_{2}/\gamma_{1}$. The atoms are more entangled when the spontaneous emission rate of the driven atom is larger than the other atom. 

We also see the concurrence plateau for $\Omega_{0}<\Delta_{0}$ whose magnitude is independent of $\Omega_{0}$ but decreasing with increasing $\Delta_{0}$. The presence of the plateau for small $\Delta_{0}$ indicates that at these frequency differences the atoms are indistinguishable. This reflects the fact that  photons emitted spontaneously by each atom induce in the other atom oscillations that are partially coherent with their own spontaneous oscillations and can thus lead to an entanglement. Thus, in order to entangle the atoms on demand, the frequency difference $\Delta_{0}$ should be quite large to prevent the atoms to exchange photons of an arbitrary frequency. In other words, the process of turning on or turning off entanglement on demand can be better controlled at large~$\Delta_{0}$.

It is worth noting that similar behave of the concurrence is found for a detuned driving, $\Delta_{L}\neq 0$ that the resonant peak appears at $\Omega =\Delta$, where $\Omega = (\Omega_{0}^{2} +\Delta_{L}^{2})^{1/2}$ is the Rabi frequency in the detuned field and $\Delta = \Delta_{0}+\Delta_{L}$. 

The numerical results have shown that the atoms decay to a stationary entangled state only when the Rabi frequency is tuned to the frequency difference between the atomic transition frequencies, i.e. when $\Omega =\Delta$. We now calculate analytically the density matrix of the system and the concurrence to explain the physical origin of this feature and to explore the importance of dissipation in the creation of the entanglement. In these calculations, we first diagonalize the Hamiltonian $H_{0}$ of the system and find that it has eigenstates 
\begin{eqnarray}
\ket 1 &=& \ket{e_1} |+\rangle ,\quad \ket 2 =  |e_1\rangle |-\rangle, \nonumber\\
\ket 3 &=& |g_1\rangle |+\rangle ,\quad \ket 4 =  |g_1\rangle  |-\rangle , \label{a6}
\end{eqnarray}
with the corresponding energies
\begin{eqnarray}
E_{1,2} = \frac{1}{2}\hbar (\Delta \pm\Omega) ,\quad E_{3,4} = -\frac{1}{2}\hbar ( \Delta\mp \Omega) ,\label{a7}
\end{eqnarray}
where 
\begin{eqnarray}
|+\rangle = \cos \theta |e_2\rangle\!+\!\sin \theta |g_2\rangle , \,
|-\rangle = \sin \theta |e_2\rangle\!-\!\cos \theta |g_2\rangle ,\label{a8} 
\end{eqnarray}
are the dressed states of the driven atom and the angle $\theta$ is determined by $\cos^2 \theta=\frac{1}{2} +\Delta_L/2\Omega$.  Note that the energy states are in the form of product states indicating that the subsystems are separable. Thus, the subsystems remain separable until spontaneous emission is included. Furthermore, the states are non-degenerate unless $\Omega =\Delta$ and then the states~$\ket 2$ and $\ket 3$ become degenerate. In the physical terms, the condition $\Omega =\Delta$ corresponds to resonance of the Rabi sideband $\omega_{L}+\Omega$ of the dressed-atom with the transition frequency of the undriven atom.

We now include spontaneous emission and rewrite the master equation in terms of dressed-atom operators $R^{-} =\ket{-}\bra{+}, R^{+} =\ket{+}\bra{-}$, and $R_{z}\!=\!\frac{1}{2}({\ket +}{\bra +} - \ket -\bra -)$, and make the unitary transformation on the density matrix of the system
\begin{eqnarray}
\tilde{\rho} = \exp(i\tilde{H}t/\hbar)\rho \, \exp(-i\tilde{H}t/\hbar) ,\label{a9}
\end{eqnarray}
where $\tilde{H} = \hbar \Delta S_1^z+ \hbar \Omega R_{z}$.  

The master equation for the transformed density operator is found to contain terms oscillating at frequencies $\Delta$ and $\Delta \pm\Omega$. When $\Omega\neq \Delta$ and the frequency difference is much larger than the atomic decay rates, $\Delta\gg \gamma_{1},\gamma_{2}$, these terms oscillate rapidly in time and, in the limit of $t\rightarrow\infty$, make negligible contribution to the master equation. After discarding them, which is a form of the secular approximation, we readily find that the master equation reduces to
\begin{eqnarray}
\dot{\tilde{\rho}} = {\cal L}_{1}\tilde{\rho} = \gamma_{1}\!\left(2S_1^-\tilde{\rho} S_1^{+} -S^{+}_{1}S^{-}_{1}\tilde{\rho} - \tilde{\rho}S^{+}_{1}S^{-}_{1}\right) ,\label{a10}
\end{eqnarray}
which contains only the term representing the damping of the undriven atom. No terms representing the other atom are involved, which explains why entanglement is not found in the system when $\Omega\neq \Delta$.

The situation changes when $\Omega =\Delta$. In this case, the terms oscillating at $\Delta +\Omega$ become non-oscillatory and the other become rapidly oscillating at frequencies $\Delta$ and $2\Delta$. When we discard these rapidly oscillating terms, we find the master equation has the form 
\begin{eqnarray}
\dot{\tilde{\rho}} = {\cal L}(\tilde{\rho}) = {\cal L}_{1}\tilde{\rho} + {\cal L}_{d}\tilde{\rho} +{\cal L}_{c}\tilde{\rho} ,\label{a11}
\end{eqnarray}
where ${\cal L}_{d}$ is an operator representing the damping of the dressed-atom system
\begin{align}
{\cal L}_{d}\tilde{\rho} &= \gamma_{0}\!\left(\left[R_{z},\tilde{\rho} R_{z}\right] +\left[R_{z}\tilde{\rho}, R_{z}\right]\right) 
 + \gamma_{+}\!\left(\left[R^{-},\tilde{\rho} R^{+}\right]\right. \nonumber\\
 &\left. +\left[R^{-}\tilde{\rho}, R^{+}\right] \right)\!+ \gamma_{-}\!\left(\left[R^{+},\tilde{\rho} R^{-}\right]\!+\!\left[R^{+}\tilde{\rho}, R^{-}\right]\right) ,\label{a12}
\end{align}
with $\gamma_{0}=(\gamma_{2}/4)\sin^{2}(2\theta), \gamma_{+}=\gamma_{2}\cos^{4}\theta, \gamma_{-}=\gamma_{2}\sin^{4}\theta$, and~${\cal L}_{c}$ is the operator representing the dissipative coupling between the undriven atom and the dressed-atom system 
\begin{align}
{\cal L}_{c}\tilde{\rho} =  \bar{\gamma}_{12}\left( \left[\tilde{\rho}S_1^+,R^-\right] +\left[S_1^+,R^- \tilde{\rho}\right]  +{\rm H.c.}\right) , \label{a13}
\end{align}
with $\bar{\gamma}_{12} =\gamma_{12}\cos^{2}\theta$.

Here, the master equation contains the dissipative terms of both subsystems. However, no coherent terms are involved, indicating the absence of coherent dynamics in the system. Nevertheless, the system can be driven to a stationary entangled state due to the presence of the dissipative coupling between the subsystems.

In order to examine the final state of the subsystems, we solve the equation ${\cal L}(\tilde{\rho})=0$ to find the steady-state density matrix of the system. It is easily verified that in the density matrix, written in the basis of the product states~(\ref{a6}), has a~$X$-state form~\cite{yue06}
\begin{align}
\rho & = \left(\begin{array}{cccc}
\rho_{11} & 0  & 0 & 0  \\
0 & \rho_{22}  & \rho_{23} & 0 \\
0 & \rho_{32} & \rho_{33} & 0  \\
0 & 0 & 0 & \rho_{44} 
\end{array}\right) ,\label{a14}
\end{align}
where the non-zero density matrix elements are
\begin{align}
\rho_{11} &= \gamma_{-}^2 \bar{\gamma}_{12}^2 /D ,\quad
\rho_{22} = \gamma_{-}\left(\gamma_{1}+\gamma_{+}\right) \bar{\gamma}_{12}^2 /D ,\nonumber\\
\rho_{44} &= 1 -\gamma_{-}\left[\gamma\gamma_{1}\left(\gamma -\gamma_{0}\right) +3\gamma_{-}\bar{\gamma}_{12}^{2}\right]/D ,\nonumber\\
\rho_{23} &= \rho_{32} = \gamma_{1} \gamma_{-}\left(\gamma -\gamma_{0}\right)\bar{\gamma}_{12}/D , \label{a15}
\end{align}
with 
\begin{align}
D &= \left(\gamma -\gamma_{0}\right)^{2}\left[\gamma_{1}\left(\gamma_{+}+\gamma_{-}\right) -\bar{\gamma}_{12}^{2}\right] \nonumber\\
& +\left(\gamma_{+}+\gamma_{-}\right)\left[\gamma_{0}\gamma_{1}\left(\gamma -\gamma_{0}\right) +4\gamma_{-}\bar{\gamma}_{12}^{2}\right] ,\label{a16}
\end{align}
and $\gamma =\gamma_{1}+\gamma_{0} +\gamma_{+} +\gamma_{-}$ is the total damping rate.

The arrangement that the laser field only couples to one of the two atoms resembles very much an open cascade system~\cite{gz00,ma00}. That is, the output (fluorescence) of the laser driven atom drives the other atom with no feeding back onto the first atom. In this case, the spontaneous emission coupling term has the form
\begin{eqnarray}
{\cal L}_{c}\tilde{\rho} = \bar{\gamma}_{12} \left(\left[S_{1}^{+},R^{-} \tilde{\rho}\right] + \left[\tilde{\rho}R^{+},S_{1}^{-}\right]\right) ,\label{a17}
\end{eqnarray}
and then it is straightforward to show that the steady-state density matrix elements are very similar in form to the ones found for the mutually driving systems. They differ only in that the denominator $D$ is replaced by 
\begin{align}
D^{\prime} = \left(\gamma_{+}+\gamma_{-}\right)\left[\gamma\gamma_{1}\left(\gamma -\gamma_{0}\right) +2\gamma_{-}\bar{\gamma}_{12}^{2}\right] .\label{a18}
\end{align}  

\begin{figure}[htpb!]
\includegraphics[width=0.45\columnwidth]{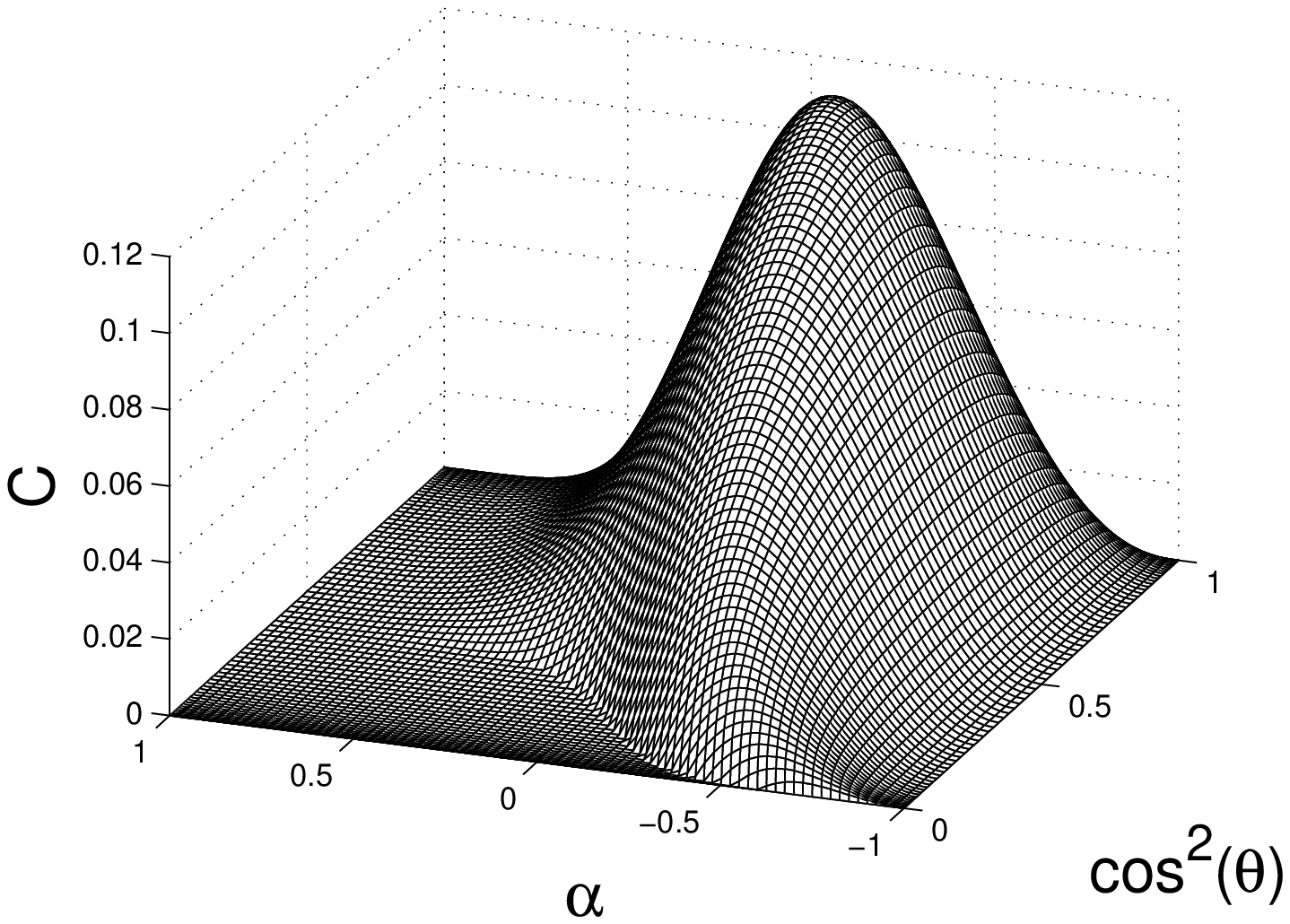}
\includegraphics[width=0.45\columnwidth]{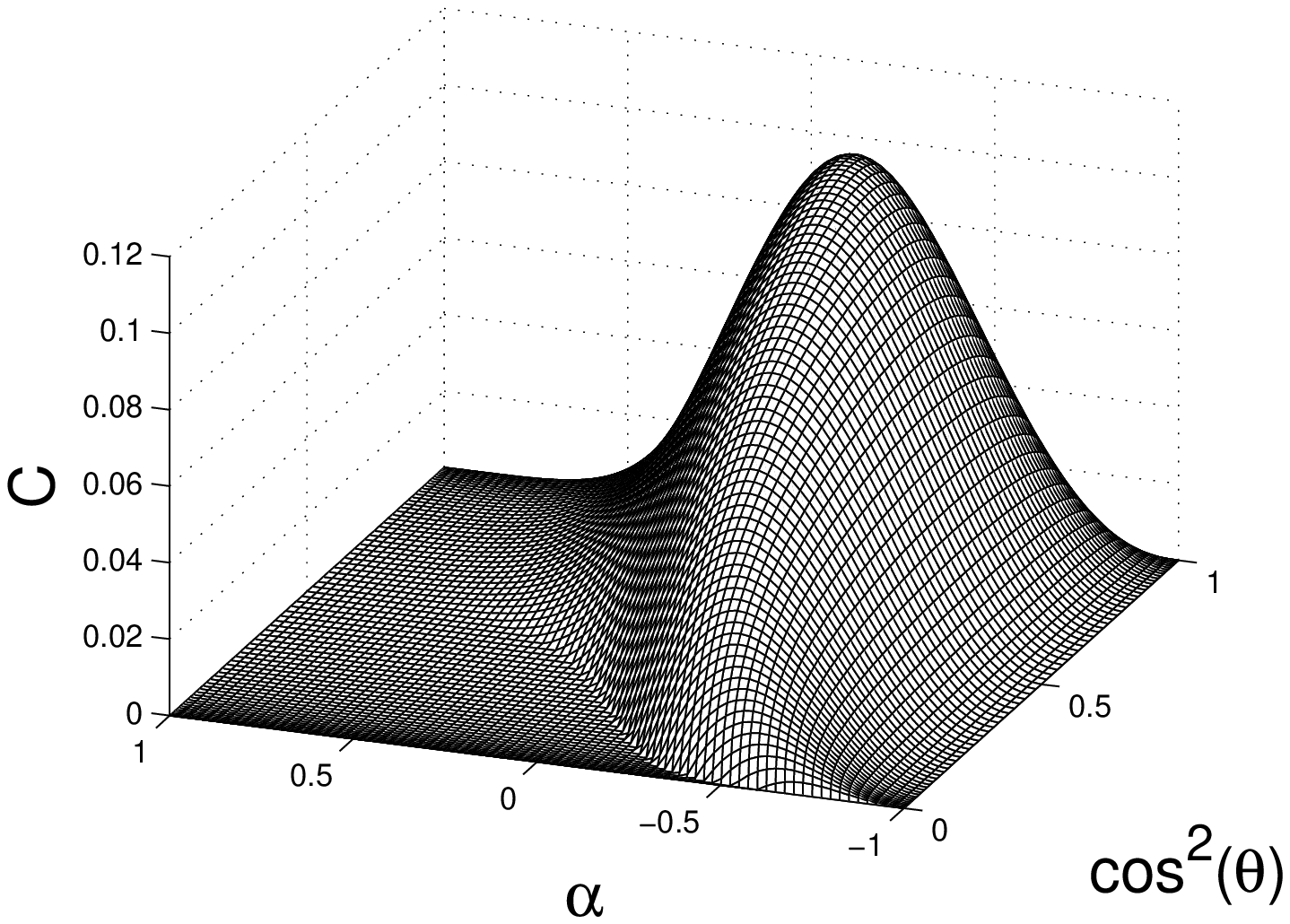} 
\caption{{\footnotesize Steady-state concurrence as a function of $\alpha =(\gamma_{1}-\gamma_{2})/(\gamma_{1}+\gamma_{2})$ and $\cos^{2}\theta$ for mutually driving subsystems (left frame) and for the case of a cascade open system (right frame).}}
\label{fig2}
\end{figure}

The $X$-state form of the density matrix allows us to obtain a simple formula for the concurrence   quantifying the amount of entanglement between the undriven and the dressed-atom systems at the level crossing point of $\Omega =\Delta$.  
The concurrence is shown in Fig.~\ref{fig2} for continuously varying $\gamma_{2}/\gamma_{1}$ and~$\Delta_{L}$. The left frame illustrates the concurrence for the case when both subsystems react to the mutually emitted photons, whereas the left frame illustrates the concurrence for the case of an open cascade system, where the undriven atom reacts to the photons emitted by the dressed-atom system, while there is no interaction in the reverse direction. From the data, we note that the concurrence reaches a maximum height for a large positive laser detuning, $\Delta_{L}>0$, and $\gamma_{2}>\gamma_{1}$. In terms of the population of the dressed-states of the driven atom, the condition $\Delta_{L}>0$ corresponds to no inversion on the dressed-atom transition, the higher frequency Rabi sideband, which for $\Omega =\Delta$ is resonant to the transition frequency of the undriven atom. The atoms remain separable for practically all $\gamma_{1}>\gamma_{2}$. Thus, the driven atom must decay faster than the undriven one to entangle the atoms by dissipation. Needless to say, the fluorescence from the driven atom acts as a pump for entanglement between the atoms even if both systems react to the mutually emitted photons. 

The shape of the concurrence for the cascade open system, illustrated in the right frame of Fig.~\ref{fig2}, continues the trend of the case of the mutually driving systems, shown in the left frame. However, it should be pointed out that the analogy between the two cases is not absolute because there is always a non-zero probability that the driven atom will absorb a photon emitted by the undriven atom. Despite this, Fig.~\ref{fig2} shows that the concurrence is formally almost the same for both cases and therefore the system can be regarded as an open cascade system.
 
A non-zero value of the concurrence reveals the existence of entangled states in the system. These can be easily found diagonalizing the density matrix~(\ref{a14}), which results in superposition (entangled) states
\begin{align}
\ket {s} = \cos\phi\ket{2} + \sin\phi\ket{3} ,\, 
\ket {a} = \sin\phi\ket{2} - \cos\phi\ket{3} , \label{a19}
\end{align}
where the angle $\phi$ is determined by $\cos^{2}\phi =\frac{1}{2} +\delta/(2G)$, with $\delta=\rho_{22}-\rho_{33}$ and $G=\sqrt{\delta^{2} +4|\rho_{23}|^{2}}$.
Clearly, the stationary state of the system is a mixed state involving two entangled states. The mechanism for the creation of the entangled states is the coherence $\rho_{23}$ generated by the interaction of the atoms with a dissipative reservoir.

In summary, we have proposed a scheme for the creation of entanglement by the dissipative process of spontaneous emission. The scheme involves two non-identical two-level atoms separated at a small distance and interacting with a common reservoir. Arranging the system such that the undesirable dipole-dipole interaction between the atoms is suppressed and only one atom is addressed by an external laser field, we have found that the atoms can be entangled on demand by tuning the Rabi frequency of the laser field to the difference between the atomic transition frequencies. At this frequency, two energy levels of the system become degenerate resulting in a large coherence causing the system to evolve into a stationary entangled state. We have also found that the entanglement occurs only when the damping rate of the driven atom is larger than the other atom. 
We have demonstrated that this system predominately behaves as a cascade open system where the fluorescence from the dressed atom drives the other atom with no interaction in the reverse direction.

\end{document}